\documentclass[conference]{IEEEtran}
\IEEEoverridecommandlockouts
\usepackage{cite}
\usepackage{amsmath,amssymb,amsfonts}
\usepackage{algorithmic}
\usepackage{graphicx}
\usepackage{textcomp}
\usepackage{xcolor}
\usepackage[left=1.62cm,right=1.62cm,top=1.9cm]{geometry}
\usepackage{bigints}
\usepackage{mathtools}
\def\BibTeX{{\rm B\kern-.05em{\sc i\kern-.025em b}\kern-.08em
    T\kern-.1667em\lower.7ex\hbox{E}\kern-.125emX}}
\begin{document}

\title{Investigation into the role of the Bessel function order in the Fourier-Bessel series and the Hankel Transform\\
}
\author{Suketu P Patni, Vikram M Gadre \\ \textit{Department of Electrical Engineering} \\ \textit{Indian Institute of Technology, Bombay} \\ Email: 23b1299@iitb.ac.in, vmgadre@ee.iitb.ac.in}

\maketitle


\begin{abstract}
This work focuses on estimating the number of terms of a Fourier-Bessel series of order \(p'\) required to get within a certain error of a Bessel function of a fixed order \(p\) where \(p \neq p'\). Our approach consists of two steps: one, constructing an invariant over \(n\) of the \(n^{\text{th}}\) order Hankel transform; and two, observing the effect of expanding a suitably scaled Bessel function of a fixed order \(p\) in its Fourier-Bessel series of order \(p'\). We demonstrate a new error metric to simplify the error computations. Further, we generate an empirical model using numerical simulations and examine its capabilities in predicting the number of terms required.
\end{abstract}
\begin{IEEEkeywords}
Bessel's differential equations, Bessel function, Fourier-Bessel series, Hankel transform.
\end{IEEEkeywords}

\section{Introduction and Applications}
Real world signals need to be represented as deterministic signals to extract useful information from them. This can be done by expressing them as linear combination of orthogonal basis functions. Many practical signals like seismic, speech and image signals are highly redundant and require compact representation for economic storage and transmission bandwidth \cite{pkc1}. The frequency information in these signals is packed in fewer samples by a Fourier series representation than by time domain representation. In \cite{bhatt} it has been shown that the Fourier-Bessel representation is a good basis for expressing non-stationary signals, because:
\begin{enumerate}
    \item Non-stationary signals are represented in a better way using non-stationary bases \cite{clark}. Since Fourier-Bessel series use non-stationary Bessel functions as their basis set, it can better represent non-stationary signals.
    \item Wideband signals are more compactly represented as a Fourier-Bessel series expansion \cite{fb_review}.
\end{enumerate}
Fourier-Bessel series use Bessel functions of the first kind in the same way that the Fourier series uses sinusoids and cosinusoids. For aperiodic signals, the Hankel transform presents an analog for the Fourier transform, using Bessel functions instead of phasors. The Hankel transform is equivalent to the Fourier transform written in hyperspherical coordinates \cite{avery}, so in multidimensional signals with radial symmetry, it has an obvious advantage.
\par Fourier-Bessel series have been applied to many signal processing problems. Gurgen and Chen \cite{speech} approximated noise in transmitted signals as a Fourier-Bessel series. This approximation could be subtracted from the signal to enhance speech synthesis, with the Fourier-Bessel series acting as a low pass filter. Suresh et al. \cite{micro_doppler} used Fourier-Bessel series in conjunction with fractional Fourier transforms to decompose data collected by radars to capture micro-Doppler target signatures. The micro-Doppler signatures could then be analyzed to estimate the motion of detected targets. He et al. \cite{freq_modulation} used Fourier-Bessel series for signal decomposition from various sources. The decomposition process helped to identify the instantaneous frequency of the sources, which is an important parameter to separate sources from each other and from the recording frequency. However, even very recent work on Fourier-Bessel series and Hankel Transforms like in \cite{uncertainty} is largely silent on the role of Bessel function order in the series and the Transforms.
\par In this work we introduce a new metric to compute the error between a Bessel function and its Fourier-Bessel series. We provide code for numerical simulation of the behaviour of this error. We also construct a new invariant of the Hankel transform. 
\section{Notation and Basics}
\subsection{Bessel functions}
The Bessel functions are an important class of functions satisfying Bessel's differential equation
\begin{equation}
x^2 \frac{d^2 y}{dx^2} + x \frac{dy}{dx} + (x^2 - n^2)y = 0 \label{bessel-diff-eq}
\end{equation}
where $n \in \mathbb{C}$ is called the order of the Bessel function. The above equation allows for two canonical linearly independent solutions, $J_n(x)$ and $Y_n(x)$. In this paper we deal exclusively with $J_n(x)$, called the \textbf{Bessel function of the first kind of order $n$}. It is well-behaved and smooth (i.e. infinitely differentiable) over $\mathbb{R}^{+} \cup \{0\}$.

\subsection{Fourier-Bessel series expansion}
Given a fixed order $\alpha$, our basis for the vector space $V$ of smooth, real-valued functions $f$ with domain $[0, R], R \in \mathbb{R}^+$ such that $f(R) = 0$ over the field $(\mathbb{C}, +, \cdot)$ is the sequence
\[B = \left\{J_\alpha\left(\frac{u_{\alpha, n}x}{R}\right)\right\}_{n \in \mathbb{N}}\]
Throughout the paper, $u_{p, q}$ will denote the $q^{\text{th}}$ positive root of $J_p(x)$. How we express a vector $x(t) \in V$ using this basis is given by its Fourier-Bessel series:-
\vspace{0.1em}
\begin{equation}
x(t) = \sum\limits^{\infty}_{n=1} c_n J_{\alpha}\left(\frac{u_{\alpha, n}t}{R}\right) \label{fbse}
\end{equation}
where 
\begin{equation}
c_n = \frac{2 \bigintssss\limits^{R}_{0} rx(r)J_{\alpha}\left(\frac{u_{\alpha, n}r}{R}\right)dr}{\left(RJ_{\alpha + 1} (u_{\alpha, n})\right)^2} \label{fbse-coeff}
\end{equation}
This expression for $c_n$ can be derived trivially by considering that the binary operation on $V$ defined as
\begin{equation} 
\langle f, g \rangle \coloneq \bigintssss\limits^{R}_{0} xf(x)g(x)dx \label{inner-prod}
\end{equation}
is an inner product, and that
\begin{align}
&\left\langle J_\alpha\left(\frac{u_{\alpha, m}x}{R}\right) , J_\alpha\left(\frac{u_{\alpha, n}x}{R}\right) \right\rangle \notag \\
&= \begin{cases} 
      0, & m \neq n \\
      \frac{R^2}{2} \left(J_{\alpha+1}(u_{\alpha, m})\right)^2, & m = n
   \end{cases} \label{orthogonality}
\end{align}
So $B$ forms an orthogonal (albeit not orthonormal) basis over $V$ (note that the above relation holds only for $\alpha > -1$, so we will enforce this condition hereafter).

\subsection{Hankel Transform}
Let $n \in [-1/2, \infty)$.The $n^{\text{th}}$ order Hankel transform is the continuous analog for the Fourier-Bessel series, and is given by 
\begin{equation}
H_n(f(r)) = F_n(\alpha) = \int\limits^{\infty}_{0} f(r) J_n(\alpha r) r dr \label{hankel-transform}
\end{equation}
for some function $f(r)$ (provided that the integral exists), and its inverse is 
\begin{equation}
H^{-1}_n(F_n(\alpha)) = f(r) = \int\limits^{\infty}_{0} F_n(\alpha) J_n(\alpha r) \alpha d\alpha \label{inverse-ht}
\end{equation}

\section{The invariant}
We try to get some arbitrary function $f(t)$ in a similar form as the LHS of \eqref{bessel-diff-eq}. So, in \eqref{inverse-ht},
\begin{equation}
f'(r) = \int\limits^{\infty}_{0} \alpha^2 F_n(\alpha) J'_n(\alpha r) d \alpha
\end{equation}
and
\begin{equation}
f''(r) = \int\limits^{\infty}_{0} \alpha^3 F_n(\alpha) J''_n(\alpha r) d \alpha
\end{equation}
Scale and add as 
\[r^2 f''(r) + rf'(r) = \int\limits^{\infty}_{0} \alpha F_n(\alpha) (\alpha^2 r^2 J''_n(\alpha r) + \alpha r J'_n(\alpha r)) d \alpha\]
But from \eqref{bessel-diff-eq}, 
\[\alpha^2 r^2 J''_n(\alpha r) + \alpha r J'_n(\alpha r) = (n^2 - \alpha^2 r^2) J_n(\alpha r)\]
So
\begin{align}
r^2 f''(r) + rf'(r) &= \int\limits^{\infty}_{0} \alpha F_n(\alpha) (n^2 - \alpha^2 r^2) J_n(\alpha r) d \alpha \notag \\
&= n^2 f(r) - r^2 \int\limits^{\infty}_{0} \alpha^3 F_n(\alpha) J_n(\alpha r) d \alpha \notag \\
&= n^2 f(r) - r^2 H^{-1}_{n} (\alpha^2 F_n(\alpha))
\end{align}
Note that the LHS does not depend on $n$, but the RHS does. Consider that 
\begin{align}
n^2 f(r) &- r^2 H^{-1}_{n} (\alpha^2 F_n(\alpha)) \notag \\
= (n+1)^2 f(r) &- r^2 H^{-1}_{n+1} (\alpha^2 F_{n+1}(\alpha)) \notag
\end{align}
\begin{equation}
\Rightarrow \frac{f(r)}{r^2} = \frac{H^{-1}_{n+1}(\alpha^2 F_{n+1}(\alpha)) - H^{-1}_{n}(\alpha^2 F_{n}(\alpha))}{2n+1} \label{invariant}
\end{equation}
The RHS of \eqref{invariant} is our invariant. For instance, say $f(r) = r^s$ where $s \in \mathbb{R}$ and $f:(0, \infty) \to \mathbb{R}$. We have the following identity \cite{papoulis}:-
\begin{equation}
H_n(r^s) = F_n(\alpha) = \frac{2^{s+1} \Gamma\left(\frac{1}{2} (n + s + 2)\right)}{\alpha^{s+2}\Gamma\left(\frac{1}{2} (n - s)\right)}
\end{equation}
and since we know from \eqref{inverse-ht} that the Hankel transform is involutary, 
\begin{equation}
H^{-1}_{n}(\alpha^k) = \frac{2^{k+1} \Gamma\left(\frac{1}{2} (n + k + 2)\right)}{r^{k+2}\Gamma\left(\frac{1}{2} (n - k)\right)} 
\end{equation}
A useful result is that
\begin{align}
H^{-1}_{n}(\alpha^2 F_n(\alpha)) &=  H^{-1}_{n}\left(\frac{2^{s+1} \Gamma\left(\frac{1}{2} (n + s + 2)\right)}{\alpha^{s}\Gamma\left(\frac{1}{2} (n - s)\right))}\right) \notag \\ 
&= \frac{2^{s+1} \Gamma\left(\frac{1}{2} (n + s + 2)\right)}{\Gamma\left(\frac{1}{2} (n - s)\right)} H^{-1}_{n}(\alpha^{-s}) \notag \\
&= 4r^{s-2} \frac{\Gamma(\frac{n}{2} + \frac{s}{2} + 1)}{\Gamma(\frac{n}{2} + \frac{s}{2})} \frac{\Gamma(\frac{n}{2} - \frac{s}{2} + 1)}{\Gamma(\frac{n}{2} - \frac{s}{2})} \notag \\
&= 4r^{s-2} \frac{(n+s)}{2} \frac{(n-s)}{2} \notag \\ 
&= r^{s-2} (n^2 - s^2)
\end{align}
So the RHS of \eqref{invariant} is
\[\frac{r^{s-2}((n+1)^2 - s^2) - r^{s-2}(n^2 - s^2)}{2n+1} = \frac{r^s}{r^2}\]
Our invariant holds.
\section{Approximating Bessel functions}
Consider $x(t) = J_p\left(\frac{u_{p, q}t}{R}\right)$, with domain $[0, R]$. Here $R \in \mathbb{R}^{+}$. Clearly, $x(t) \in V$, so, we may expand it in its Fourier-Bessel series in terms of $J_{p'}$ where $p' \neq p$ as in \eqref{fbse} as
\begin{equation}
x(t) = \sum\limits^{\infty}_{n=1} c_n J_{p'}\left(\frac{u_{p', n}t}{R}\right)
\end{equation}
Consider only the first $l$ terms of this expansion i.e. consider the function 
\[x_l(t) = \sum\limits^{l}_{n=1} c_n J_{p'}\left(\frac{u_{p', n}t}{R}\right)\]
Our objective is to try and determine the minimal $l \in \mathbb{N}$ for which, given an $\epsilon > 0$, the following inequality holds :-
\[||x(t) - x_l(t)||_d < \epsilon\]
where $d$ is some norm.
\section{A new norm}
Let us see what happens if we use the usual $L^m$-norm. We need
\[||x(t) - x_l(t)||_{L^m} < \epsilon \]
\[\Rightarrow \int\limits^{R}_{0} \left|J_p\left(\frac{u_{p, q}t}{R}\right) - \sum\limits^{l}_{n=1} c_n J_{p'}\left(\frac{u_{p', n}t}{R}\right)\right|^m dt < \epsilon^m\]
For odd $m$ we are not even able to remove the absolute value sign. Further for $m$ even, say $m=2$, we require
\begin{equation}
\int\limits^{R}_{0} \left(J_p\left(\frac{u_{p, q}t}{R}\right) - \sum\limits^{l}_{n=1} c_n J_{p'}\left(\frac{u_{p', n}t}{R}\right)\right)^2 dt < \epsilon^2
\end{equation}
The integral 
\[\int\limits^{R}_{0} \left(J_p\left(\frac{u_{p, q}t}{R}\right)\right)^2 dt\]
does not seem to yield itself to any nice expressions (although it can be expressed using hypergeometric functions \cite{hypergeo}). However, if the integrand had just one more factor of $t$, we would have a nice expression for it (as given in \eqref{orthogonality}).\\\\
\textbf{\textit{Theorem 1:}} The function $||\cdot||_{L^{2'}}$ defined as
\[||x(t)||_{L^{2'}} \coloneq \left(\int\limits^{R}_{0} (x(t))^2 tdt\right)^{1/2}\]
is a norm over $V$.\\\\
\textbf{\textit{Proof:}}  Recall that a norm must satisfy non-negativity and the triangle inequality and must respect scaling under the scalars of the vector field.
Non-negativity and scaling can be easily shown. For the triangle inequality, we need to show that
\[||x(t)||_{L_{2'}} + ||y(t)||_{L_{2'}} \geq ||x(t) + y(t)||_{L_{2'}}\]
On squaring and expanding, we see that it is equivalent to showing that
\[\langle x(t), x(t)\rangle \langle y(t), y(t)\rangle \geq \langle x(t), y(t)\rangle^2\]
Notice that now we can directly use Cauchy-Schwarz, which states that for all vectors $u$, $v$ of an inner product space, $\langle u, u\rangle \langle v, v\rangle \geq \langle u, v\rangle^2$ holds. And we have already shown in \eqref{inner-prod} that $V$ is an inner product space.
\section{Using $L^{2'}$}
Now we need
\[||x(t) - x_l(t)||_{L^{2'}} < \epsilon \]
\[\Rightarrow \int\limits^{R}_{0} \left(J_p\left(\frac{u_{p, q}t}{R}\right) - \sum\limits^{l}_{n=1} c_n J_{p'}\left(\frac{u_{p', n}t}{R}\right)\right)^2 tdt < \epsilon^2 \]
Consider only the LHS. 
\begin{align*}
    = \int\limits^{R}_{0} &\left(J_p\left(\frac{u_{p, q}t}{R}\right)\right)^2 tdt + \int\limits^{R}_{0} \left(\sum\limits^{l}_{n=1} c_n J_{p'}\left(\frac{u_{p', n}t}{R}\right)\right)^2 tdt \\
    &- 2 \int\limits^{R}_{0} \left(\sum\limits^{l}_{n=1} c_n J_p\left(\frac{u_{p, q}t}{R}\right)J_{p'}\left(\frac{u_{p', n}t}{R}\right)\right) tdt\\
\end{align*}
\begin{align*}
    = \frac{R^2}{2} &\left(J_{p+1}(u_{p, q})\right)^2 + \sum\limits^{l}_{n=1} c_{n}^{2} \frac{R^2}{2} \left(J_{p'+1}(u_{p', n})\right)^2 \\
    &- 2 \int\limits^{R}_{0} \left(\sum\limits^{l}_{n=1} c_n J_p\left(\frac{u_{p, q}t}{R}\right)J_{p'}\left(\frac{u_{p', n}t}{R}\right)\right) tdt
\end{align*}
using \eqref{orthogonality}. Now consider only the last term.
\begin{align*}
    &\int\limits^{R}_{0} \left(\sum\limits^{l}_{n=1} c_n J_p\left(\frac{u_{p, q}t}{R}\right)J_{p'}\left(\frac{u_{p', n}t}{R}\right)\right) tdt \\
    &= \sum\limits^{l}_{n=1} c_n\int\limits^{R}_{0} \left(rx(r)J_{p'}\left(\frac{u_{p', n}r}{R}\right)\right) dr \\
    &= \sum\limits^{l}_{n=1} c_{n}^{2} \frac{R^2}{2} \left(J_{p'+1}(u_{p', n})\right)^2 \\
    &= \sum\limits^{l}_{n=1} \frac{R^2}{2} \left(J_{p'+1}(u_{p', n})\right)^2 \frac{4\left(\int\limits^{R}_{0} \left(rx(r)J_{p'}\left(\frac{u_{p', n}r}{R}\right)\right) dr\right)^2}{\left(RJ_{p'+1}(u_{p', n})\right)^4} \\
    &= 2R^2\sum\limits^{l}_{n=1} \frac{1}{\left(J_{p'+1}(u_{p', n})\right)^2} \left(\int\limits^{1}_{0} J_p(u_{p, q}t)J_{p'}(u_{p', n}t) tdt\right)^2
\end{align*}
where in the last step we substitute $t = \frac{r}{R}$. So the LHS turns out to be just
\begin{align}
&\frac{R^2}{2} \left(J_{p+1}(u_{p, q})\right)^2 \notag\\
&- 2R^2\sum\limits^{l}_{n=1} \frac{1}{\left(J_{p'+1}(u_{p', n})\right)^2} \left(\int\limits^{1}_{0} J_p(u_{p, q}t)J_{p'}(u_{p', n}t) tdt\right)^2 \notag
\end{align}
 Writing it out in full, given an $\epsilon > 0$, our aim is to find the minimal positive integer $l$ for which
\[\sum\limits^{l}_{n=1} \left(\frac{\bigintssss\limits^{1}_{0} J_p(u_{p, q}t)J_{p'}(u_{p', n}t) tdt}{J_{p'+1}(u_{p', n})}\right)^2 > \frac{\left(J_{p+1}(u_{p, q})\right)^2}{4} - \frac{\epsilon^2}{2R^2}\]
A small corollary is that the LHS is always positive, so if the RHS is negative i.e. if
\[\frac{(J_{p+1}(u_{p, q})^2}{4} - \frac{\epsilon^2}{2R^2} < 0 \text{ or } \frac{\epsilon}{R} > \frac{J_{p+1}(u_{p, q})}{\sqrt{2}}\]
then $l=1$ works. One term itself is sufficient to get within the required error bound.
\section{Numerical Simulation}
It would seem that our simplification has reached the limit. Let us try and resort to numerical simulation instead. Consider $p=0$ and $q=1$, and restrict to integer $p'$ for now. Given $p'$, the value of $l$ depends only on the ratio $\epsilon/R$. For different $p'$, our simulation will aim to find the least natural $l$ for which 
\begin{align}
\sum\limits^{l}_{n=1} \left(\frac{\bigintssss\limits^{1}_{0} J_0(u_{0, 1}t)J_{p'}(u_{p', n}t) tdt}{J_{p'+1}(u_{p', n})}\right)^2 &> \frac{\left(J_{1}(u_{0, 1})\right)^2}{4} - \frac{(\epsilon/R)^2}{2} \notag \\
&\approx 0.06738 - \frac{(\epsilon/R)^2}{2} \notag
\end{align}
Post the simulation, three particular cases for $\epsilon/R$ stand out on first sight: 0.01, 0.05 and 0.15.\\\\
\begin{tabular}{ |p{0.8cm}||p{0.4cm}|p{0.4cm}|p{0.4cm}|p{0.4cm}|p{0.4cm}|p{0.4cm}|p{0.4cm}|p{0.4cm}|p{0.4cm}|p{0.4cm}| }
 \hline
 \text{}&\multicolumn{9}{|c|}{$p'$} \\
 \hline
 $\epsilon/R$&1&2&3&4&5&6&7&8&9\\
 \hline
 \hline
 0.01&22&44&66&88&110&132&154&176&198\\
 \hline
 0.05&4&8&12&16&20&24&28&32&36\\
 \hline
 0.15&1&2&3&4&5&6&7&8&9\\
 \hline
\end{tabular}
$\text{ }$\\
$\text{ }$\\
The regularity is striking. Note that not all values of $\epsilon/R$ give such perfectly linear increase in the number of terms. A ratio of 0.12 makes the program return $(2, 3, 4, 6, 7, 8, 10, 11, 13)$ respectively for $p = 1,2,\ldots,9$. But still, this is more than enough evidence to hypothesize: \textbf{the number of terms is roughly linear in $p'$}. \\
We will now vary $\epsilon/R$ between 0.01 and 0.36 (as that is near the upper bound for $l > 1$) in steps of 0.01. For each such value, we will use a very simple linear regressor to determine the values of slope $m$ and y-intercept $c$ for the set of $l$ values that we obtain for $p' \in \{1, 2, 3, \ldots, 10\}$.
\begin{figure}[h!]
\minipage{0.5\textwidth}
    \centering
  \includegraphics[width=0.85\linewidth]{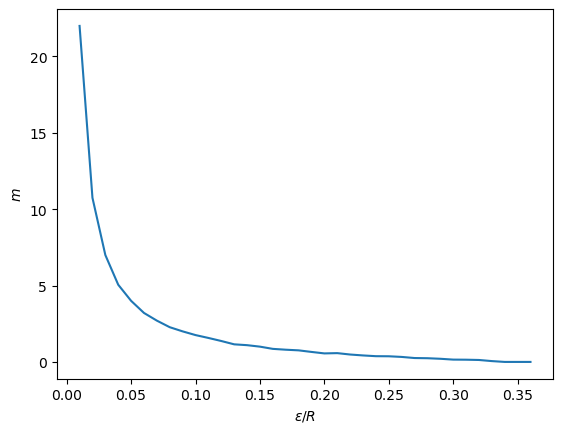}
  \caption{Slope of best fit line vs $\epsilon/R\text{ }$}\label{fig:1}
\endminipage
\end{figure}
\begin{figure}
\minipage{0.5\textwidth}
    \centering
  \includegraphics[width=0.85\linewidth]{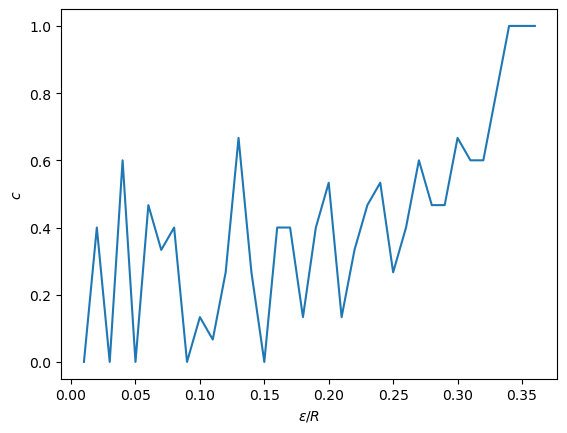}
  \caption{Intercept of best fit line vs $\epsilon/R\text{ }$}\label{fig:2}
\endminipage
\end{figure}

The slope plot, unlike the intercept plot, is very clean and useful. Since the shape resembles that of the plot of $f(x) = 1/x$, we fit that curve to an equation of the form $f(x) = a/x + b$ and get parameters $a \approx 0.2259, b \approx -0.55585$. This model satisfies an R-squared coefficient of more than 99.9\% .\\
So the general method of finding (approximately) the number of terms given $p', \epsilon/R$ is to 
\begin{enumerate}
    \item \textbf{Determine the slope using our best fit-equation, say $m_0$.}
    \item \textbf{Look up $l$ for $p'=1$ and given $\epsilon/R$, say $l_0$}, and then 
    \item \textbf{Claim that the minimal $l$ is roughly $m_0 (p' - 1) + l_0$}
\end{enumerate} 
We put this to the test. Let us find (``theoretically" and empirically) the number of terms for $\epsilon/R = 0.12$ and $p'=30$. 
\[l_0 = 2, m_0 = 0.2259/0.12 + (-0.55585) \approx 1.32665\]
\[m_0 (p'-1)+l_0 = 1.32665\times29 + 2 \approx 40.47\]
and the actual number of terms is 41. The \textbf{agreement is excellent}. We test this further by generating 4 more plots now for 4 different values of $\epsilon/R$ from 0.05 and 0.33 (say 0.05, 0.12, 0.19, 0.26). In each plot, we will plot two curves vs $p' \in [11, 20]$: one, of the empirical $l$ values, and second, of the predicted $l$ values. They are given below.
\begin{figure}[h!]
\minipage{0.5\textwidth}
  \includegraphics[width=0.45\linewidth]{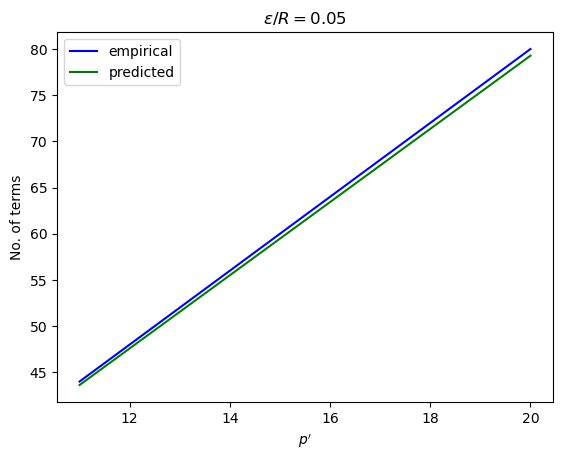}
  \includegraphics[width=0.45\linewidth]{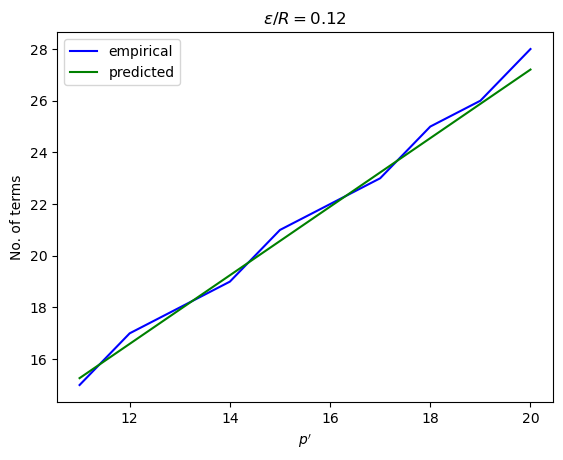}
\endminipage
\end{figure}
\begin{figure}[h!]
\minipage{0.5\textwidth}
  \includegraphics[width=0.45\linewidth]{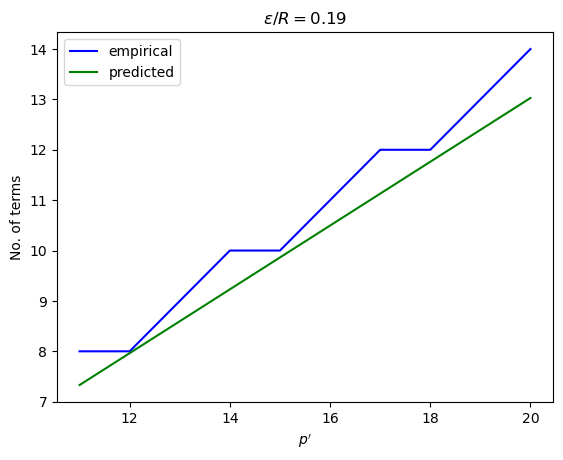}
  \includegraphics[width=0.45\linewidth]{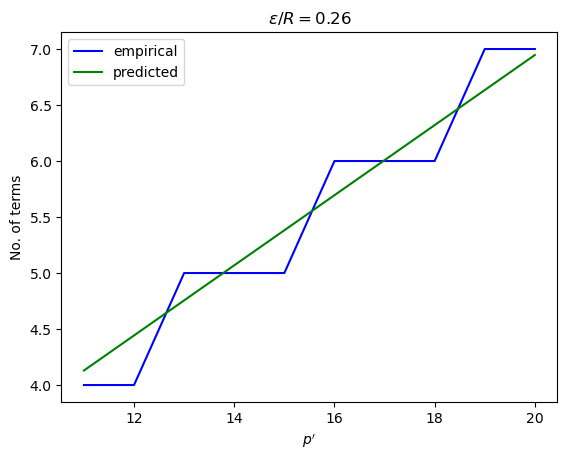}
\endminipage
\end{figure}

It is clear that \textbf{as we allow more and more error, our formula becomes less and less accurate}. Even so, our formula is \textbf{never more than an integer away from the accurate value} (note the ticks on the y-axis!), and as we allow for \textbf{less and less error, it becomes more and more accurate}.

\section{Nonzero order}

Below are some plots for a fixed error bound $\epsilon/R = 0.01$, for $p = 0, 1, 2, 3, 4, 5$ ($q$ is still 1) and for $p' = 0, 1, 2, \ldots, 10$. 
\begin{figure}[h!]
\minipage{0.5\textwidth}
  \includegraphics[width=0.45\linewidth]{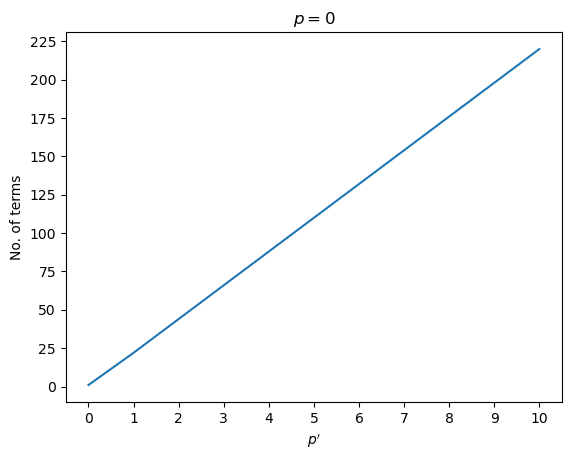}
  \includegraphics[width=0.45\linewidth]{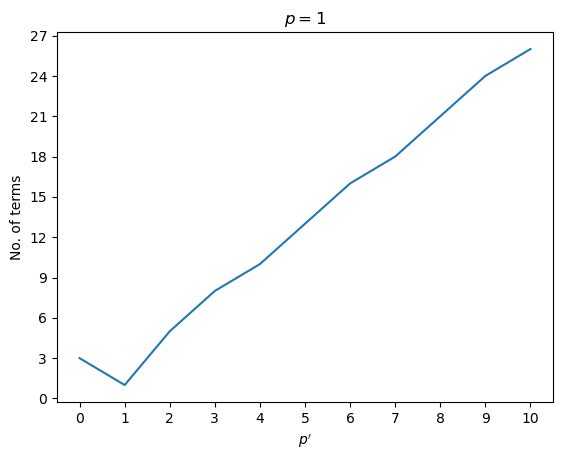}
\endminipage
\vspace{0.1em}
\minipage{0.5\textwidth}
  \includegraphics[width=0.45\linewidth]{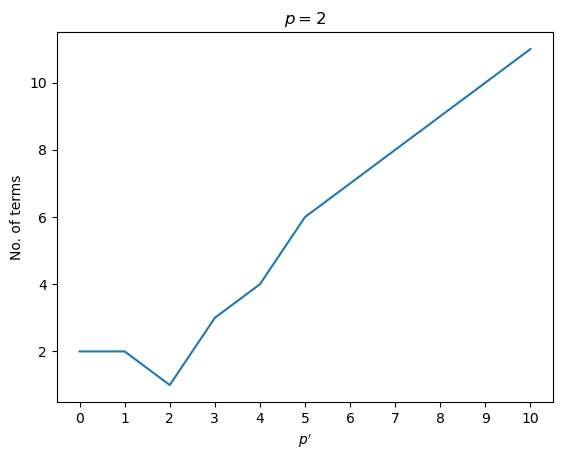}
  \includegraphics[width=0.45\linewidth]{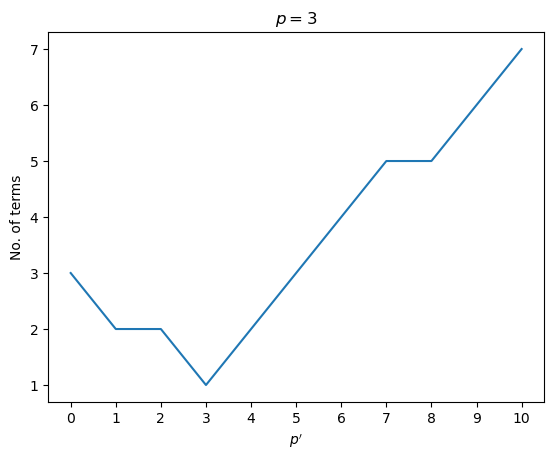}
\endminipage
\vspace{0.1em}
\minipage{0.5\textwidth}
  \includegraphics[width=0.45\linewidth]{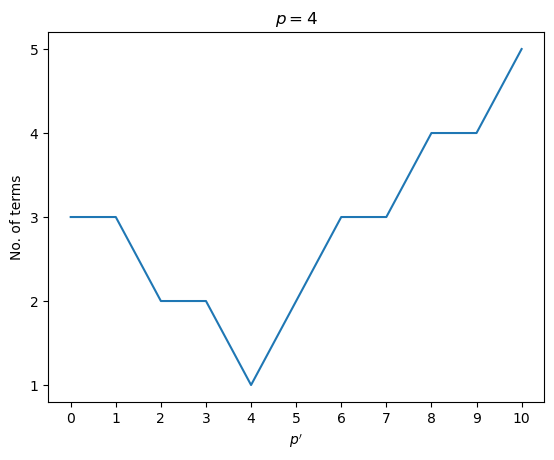}
  \includegraphics[width=0.45\linewidth]{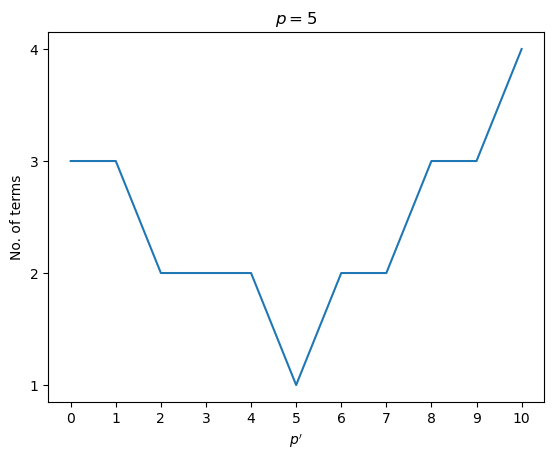}
\endminipage
\caption{Number of terms vs. Order $(q=1)$}
\end{figure}
As is expected, for the function $J_p\left(\frac{u_{p, q}t}{R}\right)$, it is in the FBSE of order $p$ that the minimum number of terms is required (in fact just one). What isn't obvious is why the number of terms is (non-strictly) monotonic on both sides of this order. As the expansion order $p'$ goes farther and farther away from $p$, the number of terms never decreases. But, look at the corresponding plots for $q=2$ below (for the same $\epsilon/R$).
\begin{figure}[h!]
\minipage{0.5\textwidth}
  \includegraphics[width=0.45\linewidth]{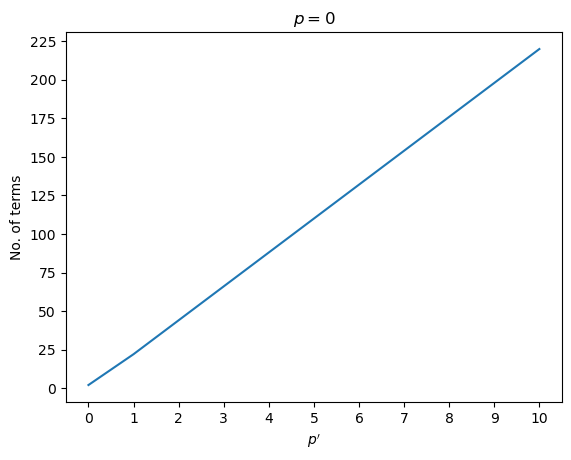}
  \includegraphics[width=0.45\linewidth]{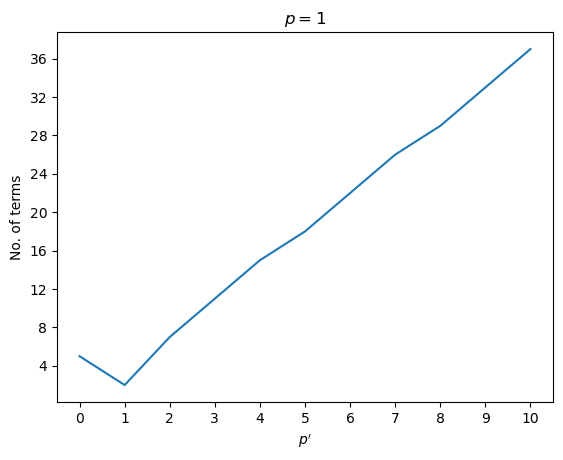}
\endminipage
\end{figure}
\begin{figure}
\vspace{0.1em}
\minipage{0.5\textwidth}
  \includegraphics[width=0.45\linewidth]{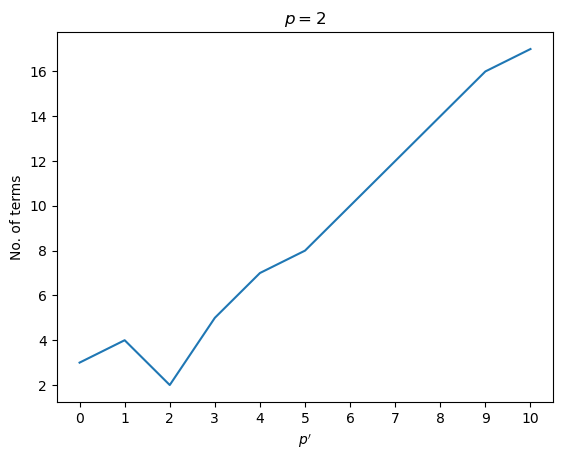}
  \includegraphics[width=0.45\linewidth]{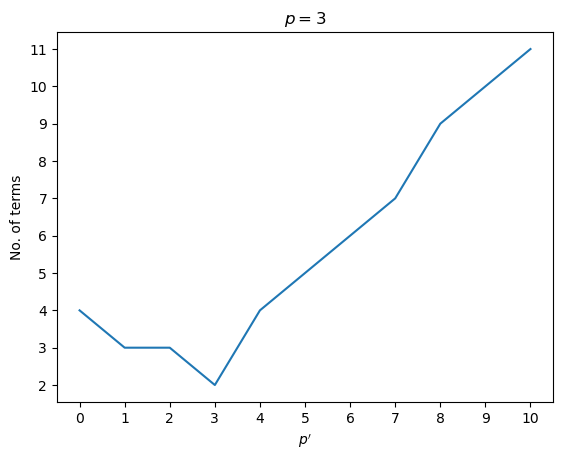}
\endminipage
\vspace{0.1em}
\minipage{0.5\textwidth}
  \includegraphics[width=0.45\linewidth]{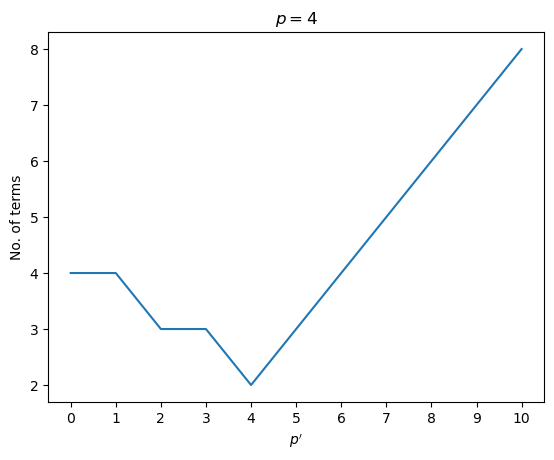}
  \includegraphics[width=0.45\linewidth]{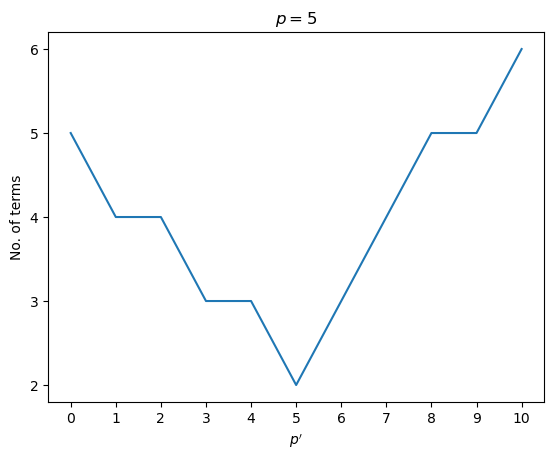}
\endminipage
\caption{Number of terms vs. Order $(q=2)$}
\end{figure}

There is a discrepancy for the $p=2$ case. Even though $p'=0$ is farther away from $p=2$, it is still a better approximator than $p'=1$. In fact this discrepancy has also been observed for $q=3$ and $q=4$. So our conjecture does not seem to hold for $q>1$.
\vspace{-0.5em}
\section{Conclusion}
Alongwith demonstrating the existence of an order-invariant for Hankel transform, we have conceived of a model to predict the number of terms required to get within a certain error of a Bessel function using a Fourier-Bessel series of different order. This shines light on the role of diversity in the Bessel functions. Further progress would involve explaining our simulation results analytically. In fact, one use case for the mentioned metric could be in computing fast approximations to signals in terms of the Bessel functions of the first kind, which would have applications in compression and transmission.
\section{Acknowledgements}
The authors would like to acknowledge the efforts of Mr. Shrikant Sharma, scientist at LRDE and PhD student in electrical engineering at IIT Bombay; P. Radhakrishna, Mr. Shrikant's research supervisor and erstwhile Director, LRDE; and Arya Vishe, B.Tech student in electrical engineering at IIT Bombay for their valuable insights and contributions to discussions relevant to the paper.
\bibliographystyle{unsrt}
\bibliography{references}

\end{document}